# Soft QCD Results from D0

**Gilvan A. Alves[1]**
*Lafex/CBPF*
*Rua Xavier Sigaud 150, Rio de Janeiro, Rj,22290-180 Brazil*
*E-mail:* `gilvan@cern.ch`

**On behalf of the D0 Collaboration**

This note presents some selected results on soft QCD studies done by the D0 collaboration at the Tevatron collider. Results on elastic proton-antiproton cross section, double parton scattering, underlying event and exclusive diffractive production are discussed, together with their implication for the LHC.

---

[1] Speaker

## 1. Introduction

In hadron collisions, the total cross section includes the sum of elastic, diffractive and hard interactions. The hard interaction itself can be divided in two components: the hard scattering process in which partons are emitted at high transverse momentum with respect to the incoming beams, and therefore can be described in perturbative QCD, and a non-perturbative soft component usually modeled by Monte Carlo event generators. An accurate understanding of the soft component is important for precision measurements of many physics processes to be studied at the LHC.

In this note we present several experimental results related to soft QCD including: elastic scattering (section 2), double parton interactions (section 3), underlying event studies (section 4) and exclusive diffractive production (section 5).

## 2. Elastic Scattering

The elastic differential cross section, $d\sigma/d|t|$, where $|t|$ is the four-momentum transferred squared, contains relevant information about the proton structure and non-perturbative aspects of proton-(anti)proton interactions. Typical elastic scattering angles are very small (less than a few milliradians), consequently (anti)protons scattered at these angles cannot be detected by the main central detectors. For Run IIa, the D0 experiment added a Forward Proton Detector (FPD)[1] to measure scattered protons and antiprotons from elastic and diffractive interactions. The data used in this analysis were taken in a dedicated store of the Tevatron collider with only one proton and one antiproton bunch colliding at the injection tune ($\beta^*=1.6$ m). The recorded luminosity was about 30 nb$^{-1}$ and a special trigger list optimized for diffractive physics was used. Figure 1 shows the measured differential cross section for the range $0.25 < |t| < 1.2$ GeV$^2$ including the best exponential fit for $0.25 < |t| < 0.6$ GeV$^2$. The extracted value for the slope parameter is $b = 16.54 \pm 0.10$ (stat.) $\pm 0.80$ (syst.) GeV$^{-2}$. The major sources of systematic uncertainty come from detector positions, efficiencies and luminosity measurement. This is the first measurement of $d\sigma/d|t|$ at $\sqrt{s} = 1.96$ TeV and also the first measurement of the diffraction minimum of the elastic differential cross section.

## 3. Double Parton Interactions

D0 has studied $\gamma$+3-jet events to measure double parton scattering (DPS), whereby two pairs of partons undergo hard interactions in a single proton antiproton collision. DPS is not only a background to many rare processes, especially at higher luminosities, but also provides insight into the spatial distribution of partons in the colliding hadrons. The DPS cross section is expressed as $\sigma_{DPS}\,\gamma$+3-jet $= \sigma_{jj}\sigma_{\gamma j}/\sigma_{eff}$, where $\sigma_{eff}$ is the effective interaction region that decreases for less uniform spatial parton distributions. D0 measures $\sigma_{eff} = 16.4 \pm 0.3$ (sta) $\pm 2.3$ (syst.) mb[2], consistent with an earlier CDF result[3], and finds $\sigma_{eff}$ to be in agreement for the three intervals (15 – 30 GeV) of jet $p_T$ in the second interaction. More precise studies can reveal a $p_T$ dependency on $\sigma_{eff}$, indicating a departure from the naive assumption that $\sigma_{DPS}$ depends on an uncorrelated product of $\sigma_{jj}$ and $\sigma_{\gamma j}$.

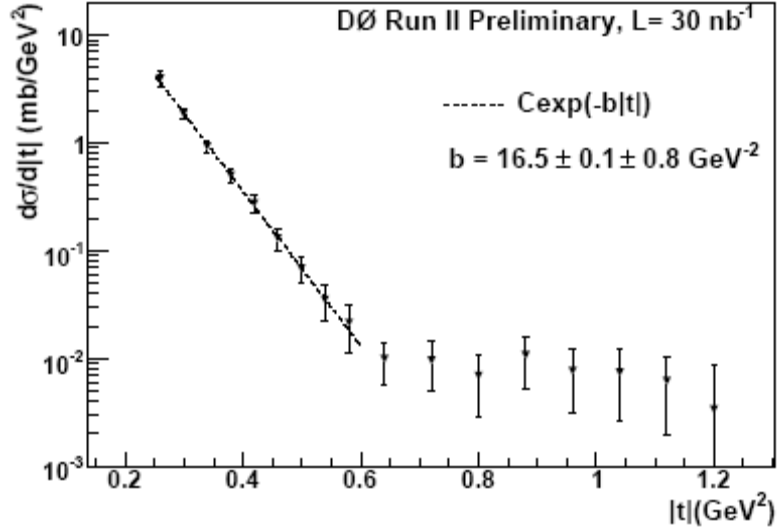

**Figure 1:** Differential cross section with its corresponding exponential fit. A normalization uncertainty of 14.3% is not shown. The uncertainties on the points are statistical and systematic added in quadrature.

## 4. Underlying Event

The underlying event (UE) consists of the beam-beam remnants minus the hard-scattering products and is becoming increasingly important to the discovery and precision potential at hadron colliders. D0 has conducted UE studies using angular distribution of tracks in Minimum Bias (MB) events[4]. The difference in azimuthal angle ($\Delta\phi$) in the detector transverse plane between the leading track and all other tracks is used to build two observables that can be compared with different Monte Carlo tunes. The data distributions were compared with three different PYTHIA tunes: Rick Field's Tune A[5], the Perugia 0 tune[6] and the Generalized Area Law model of color reconnections[7] in two $|\eta|$ ranges: $|\eta| < 1$ and $|\eta| < 2$. None of the tunes describes the data for the extended ($|\eta| < 2$) region, confirming the effectiveness of this studies for improving the tunes.

## 5. Exclusive Diffractive Production

Exclusive diffractive processes are those where the colliding hadrons emerge intact, but part of their momentum is lost producing central objects, with surrounding rapidity regions devoid of particles. D0 reported evidence for diffractive exclusive dijet production with an invariant mass greater than 100 GeV[8]. A discriminant variable ($\Delta$) based on calorimeter information was used to demonstrate a significant excess of events with very little energy outside the dijet system (Fig. 2). The probability for the observed excess to be explained by other dijet production processes is $2 \times 10^{-5}$, corresponding to 4.1 standard deviation significance. This result supports the viability of exclusive Standard Model Higgs production through p+H+p processes at the LHC, which are expected to play an important role in future studies of new physics[9].

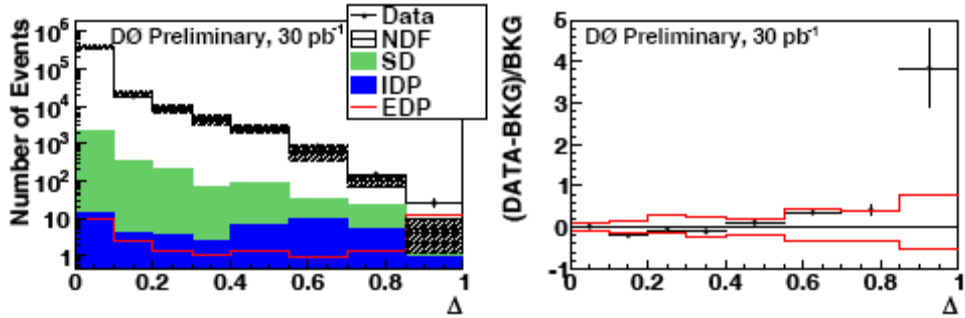

**Figure 2:** Left Figure: The discriminant Δ for data (points), exclusive diffractive signal (EDP) and background (nondiffractive (NDF), single-diffractive (SD) and inclusive double pomeron (IDP). Right Figure: The discriminant Δ for background (BKG) subtracted data divided by background. The solid red lines are ± one standard deviation systematic uncertainty on the background.

## 6. Conclusion

The D0 experiment provides the first measurement of proton-antiproton elastic cross section at $\sqrt{s}$ = 1.96 TeV. Many other soft QCD studies are underway, providing important input for understanding the interplay between soft and hard interactions necessary for the rare process searches both at Tevatron and LHC.